\documentclass[%
 reprint,
 amsmath,amssymb,
 aps,
]{revtex4-2}

\usepackage{graphicx}
\usepackage{dcolumn}
\usepackage{bm}
\usepackage{amsmath}
\usepackage{amssymb}
\usepackage{graphicx}
\usepackage{epstopdf}
\usepackage{epsfig}
\usepackage{physics}
\usepackage{hyperref}
\usepackage{lipsum}
\usepackage{graphics}

\begin{document}

\preprint{APS/123-QED}

\title{First detection of threshold crossing events under intermittent sensing}
\author{Aanjaneya Kumar}
\email{kumar.aanjaneya@students.iiserpune.ac.in}
\author{Aniket Zodage}
\email{aniket.zodage@students.iiserpune.ac.in}
\author{M. S. Santhanam}
\email{santh@iiserpune.ac.in}
\affiliation{Department of Physics, Indian Institute of Science Education and Research, Dr. Homi Bhabha Road, Pune 411008, India.}

\date{\today}

\begin{abstract}

The time of the first occurrence of a threshold crossing event in a stochastic process, known as the first passage time,
is of interest in many areas of sciences and engineering. Conventionally, there is an implicit assumption that the 
notional `sensor' monitoring the threshold crossing event is always active. In many realistic scenarios, the sensor monitoring the stochastic process
works intermittently. Then, the relevant quantity of interest is the \emph{first detection time}, which denotes the time when the sensor detects the threshold crossing event for the first time.
In this work, a birth-death process monitored by a random intermittent sensor is studied,
for which the first detection time distribution is obtained. In general, it is shown that the
first detection time is related to, and is obtainable from, the first passage time distribution. Our analytical results display an excellent agreement with simulations. Further, this framework is 
demonstrated in several applications -- the SIS compartmental and logistic models, and birth-death processes with resetting. Finally, we solve the practically relevant problem of inferring the first passage time distribution from the first detection time.\end{abstract}

\maketitle

In many situations, the time taken for an observable to reach a pre-determined threshold 
for the first time, called the first passage time (FPT), is of immense interest and 
carries practical value. The first occurrence of breakdown of an engineered structure, triggering of bio-chemical reactions, or a stock market index reaching a specific value are all examples that can be 
posed as first passage problems (FPP). This class of problems has been extensively 
studied in statistical physics for many decades \cite{redner2001,metzler_first-passage_2014,GreHolMet}, and has a large number of applications in many areas of 
physical sciences \cite{redner2001,auffinger201750,weiss1967first,szabo1980first}, engineering \cite{GreHolMet,whitmore_first-passage-time_1986,shipley_first_1972,vanvinckenroye_first_2018}, finance \cite{RemBou,zhang2009first,chicheportiche2014some} and 
biology \cite{ChouDors,iyerbiswas_first-passage_2016,zhang_first-passage_2016}. In its simplest formulation, 
the FPP is solved under an implicit assumption of perfect detection conditions -- the 
notional sensor, which monitors the first occurrence of a threshold crossing 
event, is active at all the times.

In practical applications, there is an energy cost associated with an always-on sensor, and the threshold crossing events in processes of interest are thus monitored by intermittent sensors that are not active at all times. A principal example is the wireless sensor networks widely deployed to monitor 
rare events at remote locations and operate under tight energy constraints \cite{wsn-rev-1,wsn-rev-2}. 
These sensors are typically not always active in order to optimise power consumption, and 
event sensing under such conditions can be modeled as an FPP \cite{mobilesensor, HsinLiu,intsen-robot}. Intermittent sensing has been extensively deployed in industrial and military 
environments \cite{hew2017} to detect events and is even thought to be the future trend for 
Internet-of-Things and wireless monitoring technologies \cite{HesSor}.
In several bio-chemical processes too, the sensors make stochastic transitions between active and inactive states \cite{bressloff_stochastic_2017}. A relevant example is of the heat shock response in a cell
due to environmental stresses \cite{li_rethinking_2017,sorger_heat_1991}, where the HSF family of proteins, which
upregulate heat shock proteins under heat stress, can perform upregulation only when present in their
trimeric state, while they are inactive in their monomeric state.

\begin{figure}[b]
    \centering
    \includegraphics*[width=8cm]{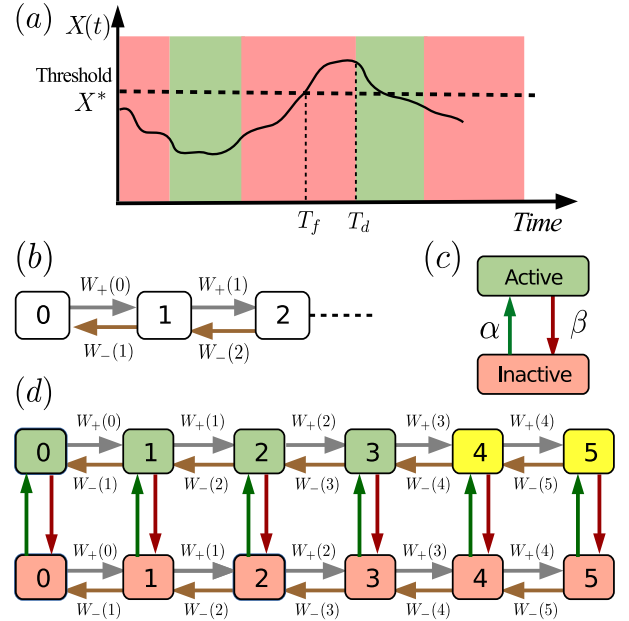}
    \caption{(a) The schematic shows the first passage time $T_f$ for a stochastic process to reach a threshold $X^*$, and the first detection time $T_d$ of the threshold crossing event under intermittent sensing. The red (green) shaded region indicates if the sensor is inactive (active). The sensor can detect events only if it is active. (b) A birth-death process models the \emph{underlying process}, (c) a sensor switching between active and inactive states. (d) The composite process with $N=5$ and $m=4$. The sensor can detect the absorbing states (yellow).}
  \label{fig1}
\end{figure}

Motivated by these phenomena, in this paper, the standard FPP for threshold crossing is expanded
to include intermittent sensing by an independent sensor that stochastically switches between 
active and inactive states. The central quantity of interest is the first detection time distribution (FDTD) of the threshold crossing event by the sensor. This can be thought of as first detection of an extreme event \cite{kishore_extreme_2011,malik_rare_2020,kumar_extreme_2020} or a general threshold
activated process \cite{greben1,bdapprox,dao_duc_threshold_2010} under intermittent sensing.
A schematic of the processes are displayed in Fig. \ref{fig1}(a), where a stochastic process 
$X(t)$ is evolving in time, while a sensor, which monitors whether the stochastic process 
has crossed a pre-defined threshold $X^*$ or not, switches between inactive (red background) 
and active (green) states. A threshold crossing event is detected only if $X(t) > X^*$ {\it and} the sensor is active at time $t$. Both the first passage time for threshold crossing and 
first detection time are marked in Fig. \ref{fig1}(a). This process is distinct from the FPP extensively studied in the context of intermittent search 
problems \cite{int-search-4,int-search-1,int-search-2,int-search-3} or intermittent target 
problems \cite{int-target-1,int-target-2,scher_unified_2021,mercado-vasquez_first_2021}. In the intermittent search/target problems, the process is terminated only when the searcher is \emph{at} the target when the searcher/target is ``active''. In contrast, in the threshold crossing under intermittent sensing, the detection of the 
crossing event can happen in any state greater than the threshold, provided the sensor 
is active (Fig.~(1)).  

We model the underlying process of interest as a Markovian, continuous-time birth-death process (BDP) 
(Fig. \ref{fig1}(b)), which has previously been used to study a variety of processes \cite{azaele_statistical_2016,novozhilov_biological_2006,crawford_transition_2012,freidenfelds_capacity_1980,ho_birthbirth-death_2018}. The state of the BDP can be interpreted as the stress or damage accumulated over time, or any other physical 
quantity where threshold crossing is of prime interest. The BDP is defined on the state space $\mathcal{S} \in \{0,1,2,\cdots,N\}$ with its dynamics 
governed by the rates  $W_{+}(j)$ and  $W_{-}(j)$ for transitioning from state $j$ to states $j+1$ and $j-1$ respectively, where $j \in \mathcal{S}$, with $W_{+}(N)=W_{-}(0)=0$. 
The probability of finding the BDP in state $n \in \mathcal{S}$ at time $t$, given that the 
system started from state $n_0\in\mathcal{S}$ initially, is given by the propagator $P_t(n|n_0)$, and its Laplace transform is denoted by $\widetilde{P}_s(n|n_0)$. A well studied quantity of the BDP is the first passage time distribution (FPTD), denoted by $F_{t}(m|n_0)$, which is probability density that the BDP reaches state $m$ for the first time, at time $t$, given that it started from state $n_0$. Through the renewal formula \cite{redner2001}, the Laplace transform of the FPTD is given by
$\widetilde{F}_{s}(m|n_0)=\frac{\widetilde{P}_s(m|n_0)}{\widetilde{P}_{s}(m|m)}$, for $n_0 \neq m$. This analysis assumes perfect detection, {\it i.e.}, as soon as the threshold $m$ is reached, the event is detected. 

To study the case of imperfect detection, an intermittent sensor is assumed and is modeled by
a two-state Markov process (Fig. \ref{fig1}(c)). The sensor dynamics is assumed to be 
independent of the underlying BDP whose threshold crossing will be monitored. The sensor can be found in any of the two states $\Omega=\{a,i\}$, with $a$ and $i$ denoting, respectively, active and inactive states. 
The sensor switches from state $i$ to $a$ at rate $\alpha$, and from $a$ to $i$ at $\beta$. For 
$\sigma_0, \sigma \in \Omega$, we define  $p_{t}(\sigma|\sigma_0)$ to be the probability that the sensor is in state $\sigma$ at time $t$, given that it was in state $\sigma_0$ initially. The Markov diagram of the composite process, {\it i.e.}, the underlying process and the sensor combined, is shown in Fig. \ref{fig1}(d) for the special case of $N=5$, and threshold $m=4$. The composite process is another continuous time Markov process on the state space $\mathbb{S}=\mathcal{S}\times\Omega$. The key object of interest is the statistics of first detection time (FDT) of a threshold crossing event, defined as the first time when the 
composite process is found in any of the states $(n,a)$ such that $n \geq m$, where $m$ is
the pre-defined threshold. Thus, the composite process has a total of $N-m+1$ absorbing states. We denote the FDTD by $D_t(n_0,\sigma_0)$, which is the probability density for the first detection event to happen at time $t$, given that the initial condition for the composite process was $(n_0,\sigma_0)$. 

In the analysis that follows, the knowledge of $P_t(n,n_0)$ for the BDP is assumed. This is known exactly for a variety of examples \citep{parthasarathy_exact_2006,sasaki_exactly_2009,crawford_transition_2012}. Furthermore, the propagator can be obtained, in principle, for any BDP governed by an $N \times N$ tridiagonal Markovian transition matrix $\mathbb{W}$ as $P_t(n|n_0) = \bra{n}e^{\mathbb{W}t}\ket{n_0}$
where $\bra{l} = ( 0 ~0 ~0 \dots 0 ~1 ~0 \dots 0)$ denotes a row vector with $1$ as its  $l^\textrm{th}$ element, with $0$ elsewhere.


To obtain the FDT statistics, we note that the FDTD satisfies the following equations
\begin{align}
    &D_t(n_0, \sigma_0) = f_1(t) + \int_0^t f_2(t') D_{t-t'}(m,i)dt',    \label{mast1} \\
    &D_t(m, i) = f_3(t) + \int_0^t e^{-\alpha t'}F_{t'}(m-1|m) D_{t-t'}(m-1,i) dt'
    \label{mast2}
\end{align}
where $n_0<m$ and the following functions are defined:
\begin{align}
 f_{1}(t) &= F_t(m|n_0) p_{t}(a|\sigma_0),~ f_{2}(t) = F_t(m|n_0)p_{t}(i|\sigma_0), \nonumber \\
 f_{3}(t) &= \alpha e^{-\alpha t} \int_t^{\infty}F_{t'}(m-1|m)dt'.
\end{align}
We obtain $D_t(m-1,i)$ from Eq.~(\ref{mast1}) in terms of $D_t(m,i)$, and taking a Laplace transform of Eqs. \ref{mast1} and \ref{mast2}, we can write
\begin{equation}
    \tilde{D}_s(n_0, \sigma_0) = \tilde{f}_1(s) + \frac{ \tilde{f}_2(s)\left(\tilde{f}_3(s)+\tilde{f}_4(s)\tilde{F}_{s+\alpha}(m-1|m)\right)}{1-\tilde{f}_5(s)\tilde{F}_{s+\alpha}(m-1|m)}.
    \label{centres}
\end{equation}
where we define:
\begin{align}
 f_{4}(t) &= F_{t}(m|m-1)p_t(a|i), ~
 f_{5}(t) = F_{t}(m|m-1)p_t(i|i).
\end{align}
Equation~\ref{centres} is our central result, that asserts that the FDTD can be obtained in terms of the FPTD alone. An alternate derivation of this result is presented in SI. The first term on the RHS, $f_1(t)$ denotes the trajectories where the FPT and FDT coincide, whereas the second term accounts for all trajectories where the first passage event goes unnoticed, and detection happens at a later time. In the limit of $\beta \to 0^+$, then $f_{2}(t) \to 0$, 
and it leads to $D_{t}(n_0,\sigma_0) = f_1(t)$. This is consistent with the expectation that in the $\beta \to 0^+$ limit deactivation of the sensor is extremely unlikely and renders the composite process 
equivalent to a simple BDP.

\begin{figure}
    \centering
    \includegraphics[width=\columnwidth]{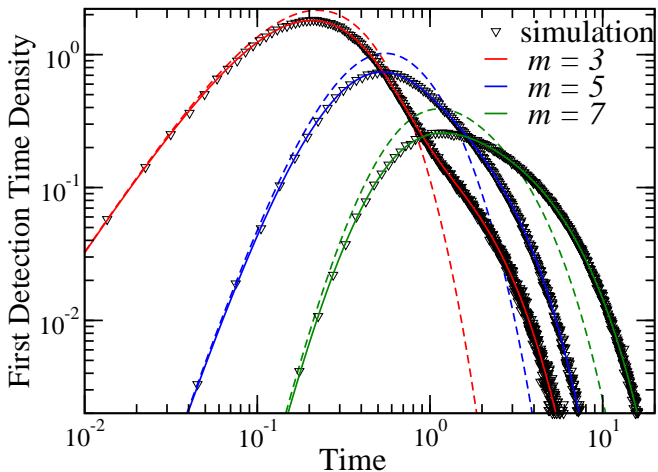}
    \caption{The FDTD for the BDP with stochastic switching (solid lines), with $n_0=0, N=20, \lambda=0.1$ and $k=1$, shown for threshold values $m = 3, 5, 7$ and $\alpha =\beta=1$. The symbols are from simulations. The dashed lines are FDTD if the sensor is {\sl not} intermittent and is always on.}
    \label{fdt1}
\end{figure}

To illustrate these results, consider a BDP with transition rates $W_{+}(j)= \lambda (N-j)$ and $W_{-}(j)= kj$, with $j \in \{0,1,2,\cdots,N\}$. These rates were previously used to model threshold crossing processes \cite{bdapprox} in the context of triggering of biochemical reactions \cite{greben1}. Figure \ref{fdt1} shows the FDTD for this process with 
parameters $n_0=0, N=20, \lambda=0.1$ and $k=1$, for threshold values $m = 3,5$ and $7$, and 
$(\alpha,\beta)=(1,1),(1,10)$ and $(10,1)$. This figure shows analytical results (solid lines) for 
which inverse Laplace transform of Eq.~\ref{centres} has been numerically performed, and 
simulations (open triangles) were generated by performing $10^7$ realizations of the stochastic process.
An excellent agreement is observed between the analytical result and the simulations.
For comparison, the case of sensor being always-on is also shown (as dashed lines) and it effectively
corresponds to the FPTD.
If the sensor is initially active, then for $t \ll \frac{1}{\beta}$, the FDTD matches with FPTD.
This is because, in this limit, the threshold is crossed earlier than the typical timescale for inactivation of the sensor. Starting from around $t \approx \frac{1}{\beta}$, the FDTD deviates from the FPTD, a feature captured by the analytical result. For $t \gg \frac{1}{\beta}$, both the FDTD and FPTD show an exponential tail, with different decay rates governed by the mean FDT and FPT respectively. 

\begin{figure}
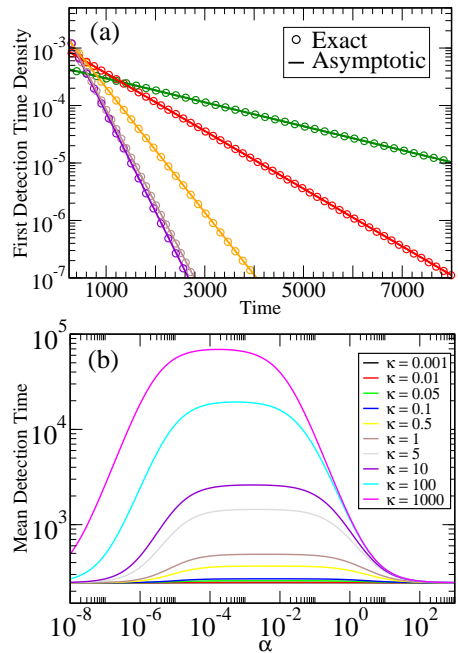

    \centering
    \includegraphics*[width=5.9cm]{fig3a.eps}
    \includegraphics*[width=5.9cm]{fig3b.eps}
    \caption{(a) Asymptotic FDTD for $t>>1$ with $m=5$ for $(\alpha,\beta)$ = $(1,1), (0.1,1), (1,0.1), (10,1)$, and $(1,10)$.. Other parameters are the same as in Fig. \ref{fdt1}. (b) Mean detection time as a function of $\alpha$, where along each curve $\frac{\beta}{\alpha}$ is held constant.}
    \label{fdt2}
\end{figure}

The mean FDT can be computed as 
\begin{equation}
    \langle T_d \rangle = \langle T_{n_0,\sigma_0} \rangle = -\frac{d}{ds} \widetilde{D}_{s}(n_0,\sigma_0) \bigg{|}_{s=0}.
\end{equation} As $t \to \infty$, the FDTD decays as $ \frac{1}{\langle T_d \rangle} e^{-t/\langle T_d \rangle}$. As shown in Fig. \ref{fdt2}(a) for a threshold of $m=5$ and for several pairs $(\alpha, \beta)$, the exact FDTD agrees with the asymptotic result for $t>>1$. Further, we define $\kappa = \beta/\alpha$, which is the fraction of time the sensor spends in the inactive state. If $\kappa << 1$, then the sensor is active nearly
all the time. The mean FDT $\langle T_d \rangle$ is plotted as a function of $\alpha$ for
constant value of $\kappa$ in Fig. \ref{fdt2}(b). As this figure reveals, the $\langle T_d \rangle \approx \langle T_f \rangle$ for small
and large values of $\alpha$. For intermediate values of $\alpha$, the mean first passage and detection
times can differ from one another by several orders of magnitude depending on the value of $\kappa$ -- larger
$\kappa$ leads to larger $\langle T_d \rangle$. This has a surprising outcome for event detections.
Physically this implies that even if $\kappa >> 1$, where the sensor spends most of its time 
in the inactive state, the detection can happen at timescales comparable to $\langle T_f \rangle$ as long as the time scales 
of the sensor switching is much faster than the intrinsic time scale of the underlying process. This 
effectively renders the switching to have little effect on the detection times. Though seemingly counterintuitive, 
a similar result was also noted in a different scenario of diffusing particles searching for intermittent 
target \citep{int-target-1}.

\begin{figure}
\includegraphics*[width=8cm]{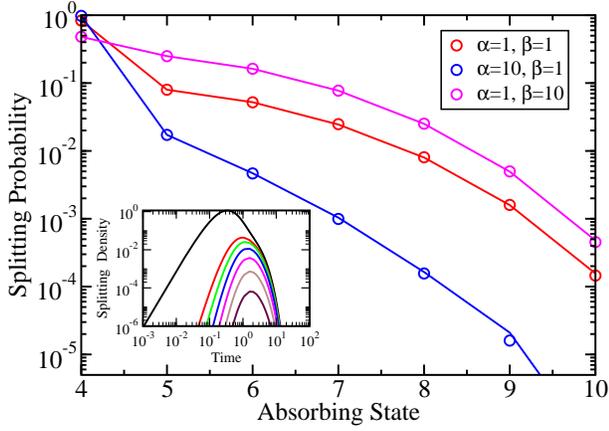}
\caption{Splitting probability for states of the underlying process (lines), for $\lambda=1$, $k=1$, $N=10$, $n_0=0$, $m=4$ for three different pairs of $(\alpha,\beta)$ = $(1,1), (10,1)$, and $(1,10)$. Symbols are from simulations. (Inset) Splitting probability density as a function of time, for $(\alpha,\beta)=(1,1)$.}
\label{splitprob}
\end{figure}

In general, the detection of the threshold crossing event does not necessarily happen at state $m$ of the underlying process, but can happen at any state $\zeta\in\{m, m+1, \cdots, N \}$. Thus, a natural question arises -- what are the splitting probabilities $H_\zeta$ that the event is detected in state $\zeta$? By performing an analysis similar to the FDTD (full calculation in SI) the density $H_t(\zeta)$ of the threshold crossing event being detected at $\zeta$ at time $t$ can be obtained. For brevity, the following quantities are defined:

\begin{align}
   \nonumber h_{1}(t) &= F(m-1|m;t) e^{-\alpha t}, \\ h_{2}(k,m, t) &= \alpha e^{-\alpha t}P_{t}(k|m,\tau_{m,m-1}>t), 
\end{align}
for $k \in \{m,m+1,\cdots,~N\}$, and $\tau_{m,m-1}$ is the first time when the underlying process visits the state $(m-1)$ starting from the state $m$. Summing over trajectories that lead to the threshold crossing event at $\zeta=m$, the Laplace transform of splitting probability density is
\begin{align}
    \widetilde{H}_{s}(m) &= \widetilde{f}_{1}(s) + \frac{\widetilde{f}_{2}(s) \left( \widetilde{h}_{1}(s) \widetilde{f}_{4}(s)+  \widetilde{h}_{2}(m,m, s)\right) }{1- \widetilde{h}_{1}(s) \widetilde{f}_{5}(s)}.
    \nonumber
\end{align}
and for $\zeta= m+r$, where $r\in \{1,2,\cdots,N-m\}$, we obtain
\begin{align}
   \widetilde{H}_{s}(m+r) = \frac{\widetilde{f}_{2}(s) \widetilde{h}_{2}(m+r,m, s)}{1 - \widetilde{h}_{1}(s) \widetilde{f}_{5}(s)}.
\end{align}
The splitting probability $H_\zeta$ can thus be obtained as $H_\zeta = \int_0^\infty dt ~ H_t(\zeta)$,
and is equal to the $s\to 0$ limit of $\widetilde{H}_{s}(\zeta)$. Figure \ref{splitprob} shows the splitting
probabilities $H_\zeta$ for different values of $\zeta$, for the parameter values $\lambda=1$, $k=1$, $N=10$, $n_0=0$, $m=4$ for three different pairs of $(\alpha,\beta)$ -- $(1,1), (10,1)$ and $(1,10)$. 
This demonstrates an excellent agreement between the analytical and simulations results. 
Furthermore, the inset in Fig.~\ref{splitprob} shows the analytically computed splitting probability density
for $(\alpha,\beta)=(1,1)$.

\begin{figure}
\includegraphics[width=8cm]{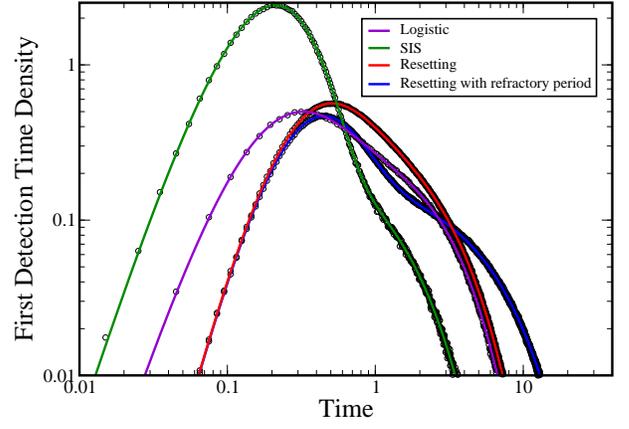}
\caption{FDTD for various processes -- Logistic model (violet), SIS epidemiological model (green), BDP with stochastic resets (red) and that with a refractory period (blue) -- showing an excellent agreement with numerical simulations of these processes (circles). Details about parameter values and additional derivations are provided in SI.}
\label{applns}
\end{figure}

If the first detection of a threshold crossing event happens at time $T_d$, can we infer the first passage time $T_f$? This has practical value as it estimates the first occurrence time for an event that possibly went
undetected. In sensors that detect abnormal voltage fluctuations (with potential for damage), $T_f$ corresponds to the time until when the device being monitored was fully functional, and $T_d$ denotes the time when sensor detects the large fluctuation. Let $F_{T_f}(m|n_0,\sigma_0, T_d)$ be the density that the first passage to threshold $m$ happens at time $T_f$, conditioned on the fact that the first detection of the threshold crossing event happens at $T_d$, and that the underlying process starts from state $n_0$ and the sensor starts from $\sigma_0$. Clearly, for $T_f>T_d$, $F_{T_f}(m|n_0,\sigma_0, T_d) = 0$. It is easy to see that for $T_f = T_d$, the probability density is $\frac{F_{T_d}(m|n_0)}{D_{T_d}(n_0,\sigma_0)}$. For $T_f<T_d$, the exact expression is
\begin{equation}
F_{T_f}(m|n_0, \sigma_0,T_d) = F_{T_f}(m|n_0) \cdot p_{T_f}(i|\sigma_0)\frac{D_{T_d-T_f}(m,i)}{D_{T_d}(n_0,\sigma_0)}.
\nonumber
\end{equation}
This shows that the FPTD conditioned on detection at a specific time $F_{T_f}(m|n_0,\sigma_0, T_d)$ can be expressed explicitly as the unconditioned FPTD $F_{T_f}(m|n_0)$, multiplied by additional \emph{tilting} factors which ensure that the threshold crossing event is detected exactly at $T_d$, after it goes undetected at $T_f$ (See SI).

\textit{Applications:} These results can be applied to threshold crossing events in processes with other absorbing states. Important examples are models of population dynamics, and compartmental models for disease propagation. These models can estimate the time taken for the size of a population, or the infected case load to cross a threshold, and such models contain an absorbing state where the size of the population, or number of infected individuals goes to zero. During a pandemic, while the dynamics of the number of infected individuals follows a continuous time BDP with disease-dependent rates, they are intermittently reported in specific time windows. Thus the formalism developed in this work has practical relevance as well. In Fig.~\ref{applns}, the analytical FDTD (details in SI) for the SIS model and the logistic model are shown along with simulation results.

These results can further be extended to processes with stochastic resetting \cite{evans_diffusion_2011,reuveni_optimal_2016,pal_first_2017,evans_stochastic_2020,pal_search_2020,pal_first_2019}, in which some observable such as the accumulated stress or damage can undergo burst-like relaxations. Recently, this process has received considerable research attention with extensive applications that include population dynamics under stochastic catastrophes \cite{di_crescenzo_note_2008,di_crescenzo_mm1_2003,kumar_transient_2000} to the dynamics of queues subject to intermittent failure. In Fig.~\ref{applns}, the FDT under intermittent sensing is shown for two cases: a BDP with simple resets, and with resets that include a refractory period \cite{evans_effects_2018}. In both cases, analytical and simulations results are in agreement. 

In this Letter, the general problem of threshold crossing under intermittent sensing is studied using the versatile BDP, and a sensor that stochastically switches between active and inactive states. A general relation between the FDTD, under intermittent sensing, and the FPTD is obtained.  The central result is that the first detection time can be obtained from the first passage times, which is known for a wide range of problems. The validity of these results is demonstrated in a variety of applications including the SIS model, logistic model, and BDP with resetting. The splitting densities, arising due to intermittency of the sensor, can be obtained by this approach. Finally, the practically relevant problem of inferring the FPTD from the FDTD is solved.

\textit{Acknowledgements.} AK gratefully acknowledges helpful discussions with Basila Moochickal Assainar and Gaurav Joshi, and the Prime Minister's Research Fellowship of the Government of India for financial support. AZ acknowledges support of the INSPIRE fellowship. MSS acknowledges the support of MATRICS grant from SERB, Govt. of India. 

\bibliography{thcross1}

\newpage
\onecolumngrid

\setcounter{page}{1}
\renewcommand{\thepage}{S\arabic{page}}
\setcounter{equation}{0}
\renewcommand{\theequation}{S\arabic{equation}}
\setcounter{figure}{0}
\renewcommand{\thefigure}{S\arabic{figure}}
\setcounter{section}{0}
\renewcommand{\thesection}{S\arabic{section}}
\setcounter{table}{0}
\renewcommand{\thetable}{S\arabic{table}}
\section{Supplementary material for ``Threshold crossing processes under intermittent sensing''}


\maketitle

\onecolumngrid


This Supplemental Material provides additional discussions and mathematical derivations which support the results described in the Letter and provides details of the various examples used to demonstrate the validity and applicability of the results.

\subsection{Interpreting the FDTD formula}

We noted in Eq.~\ref{centres}, that the first detection time distribution can be obtained from the first passage time distributions alone, which can be computed through several standard results. In Laplace space, the first detection time is expressed as
\begin{equation}
    \tilde{D}_s(n_0, \sigma_0) = \underbrace{\tilde{f}_1(s)}_{\substack{\text{First detection}\\ \text{at first passage}}} + \underbrace{\frac{ \tilde{f}_2(s)\left(\tilde{f}_3(s)+\tilde{f}_4(s)\tilde{F}_{s+\alpha}(m-1|m)\right)}{1-\tilde{f}_5(s)\tilde{F}_{s+\alpha}(m-1|m)}}_{\substack{\text{First detection happening}\\ \text{strictly after first passage event}}}.
\end{equation}
where we define:
\begin{align}
 f_{1}(t) &= F_t(m|n_0) p_{t}(a|\sigma_0),\\
 f_{2}(t) &= F_t(m|n_0)p_{t}(i|\sigma_0), \\
 f_{3}(t) &= \alpha e^{-\alpha t}\int_t^{\infty}F_{t'}(m-1|m)dt'\\
 f_{4}(t) &= F_{t}(m|m-1)p_t(a|i)\\
 f_{5}(t) &= F_{t}(m|m-1)p_t(i|i).
\end{align}
The second term on the RHS can be understood can be intuitively if expressed as the following
\begin{equation}
     \frac{ \tilde{f}_2(s)\left(\tilde{f}_3(s)+\tilde{f}_4(s)\tilde{F}_{s+\alpha}(m-1|m)\right)}{1-\tilde{f}_5(s)\tilde{F}_{s+\alpha}(m-1|m)} = \underbrace{ \tilde{f}_2(s)}_{\substack{ \text{Factor I:} \\ \text{First passage }\\ \text{while sensor} \\ \text{is inactive}}}\cdot \underbrace{ \frac{1}{1 -\tilde{f}_5(s)\tilde{F}_{s+\alpha}(m-1|m) }}_{\substack{\text{Factor II: Accounts for the} \\ \text{number of undetected threshold }\\ \text{crossings before first detection}}}\cdot\underbrace{ \left(\tilde{f}_3(s)+\tilde{f}_4(s)\tilde{F}_{s+\alpha}(m-1|m)\right)}_{\substack{\text{Factor III: Ensures detection}\\ \text{of threshold crossing event}}}
\end{equation}
Factor II is the sum of the following geometric series in Laplace space, which accounts for the number of threshold crossings that go undetected before eventual detection:
\begin{equation}
    \frac{1}{1 -\tilde{f}_5(s)\tilde{F}_{s+\alpha}(m-1|m) } =   1 + \left( \tilde{f}_5(s)\tilde{F}_{s+\alpha}(m-1|m)\right)+\left( \tilde{f}_5(s)\tilde{F}_{s+\alpha}(m-1|m)\right)^2 + \left( \tilde{f}_5(s)\tilde{F}_{s+\alpha}(m-1|m)\right)^3 + \cdots 
\end{equation}
Factor III consists of two different terms:
\begin{enumerate}
    \item $\tilde{f}_3(s)$: since the last undetected threshold crossing, the birth-death process stays above the threshold, and at time $t$, the sensor becomes active, and thus the threshold crossing event is detected.
    \item $\tilde{f}_4(s)\tilde{F}_{s+\alpha}(m-1|m)$: since the last undetected threshold crossing, the birth-death process stays above the threshold for some time and remains undetected. It then comes below the threshold, and finally, the birth-death processes reaches the threshold at time $t$ when the sensor is active.
\end{enumerate}
Both the terms add up to give two different ways of detection of the threshold crossing event, without any undetected transitions from the state $m-1$ to $m$. Overall, our central result computes the sum of probabilities of all trajectories where first detection of the threshold crossing event occurs at time $t$, and the sum can be expressed completely in terms of the first passage probabilities without any imperfect sensing.

\subsection{Alternate derivation: Computing the survival probability for a birth-death process under intermittent sensing}

In this section, we provide a step by step derivation of the first detection time statistics. Given that the dynamics of the birth death process and the sensor are independent of each other, it is instructive to first understand the dynamics of the sensor. The switching between the two states, active ($a$) and inactive ($i$), is modelled by a two-state Markov process with rates $\alpha$ for activation and $\beta$ for deactivation. For $\sigma,\sigma_0 \in \{a,i \}$, we define $p_{t}(\sigma|\sigma_0)$ as the probability that the sensor is in state $\sigma$ at time $t$, given that it was in state $\sigma_0$ initially. It is easy to obtain the following set of equations that govern the dynamics of the sensor
\begin{align}
p_t(a|a) &= \frac{\alpha}{\alpha + \beta} + \frac{\beta}{\alpha + \beta}  e^{-(\alpha + \beta)t}\\
p_t(i|a) &= \frac{\beta}{\alpha + \beta} - \frac{\beta}{\alpha + \beta}  e^{-(\alpha + \beta)t}\\
p_t(a|i) &= \frac{\alpha}{\alpha + \beta} - \frac{\alpha}{\alpha + \beta} e^{-(\alpha + \beta)t}\\
p_t(i|i) &= \frac{\beta}{\alpha + \beta} + \frac{\alpha}{\alpha + \beta} e^{-(\alpha + \beta)t}
\end{align}

Now, we shift our discussion to the computation of the first detection time distribution when the sensor is intermittent. To do this, we will adopt the approach of computing the survival probability $S_{t}(\sigma_0, n_0)$ which gives us the probability that the threshold crossing event has \emph{not} been detected until time $t$, given that, initially, the sensor was in state $\sigma_0$ and the birth death process was in state $n_0$. The Laplace transform of the FDTD and survival probability are related as $\widetilde{D}_{s}(n_0,\sigma_0) = 1 - s\widetilde{S}_{s}(n_0,\sigma_0)$.

 Let us consider the case $\sigma_0=a$.  We break up $S_{t}(a, n_0)$ in two exclusive but exhaustive parts, and write them separately. The first part contains the possible \emph{surviving} trajectories of our process such that, at time $t$, the underlying birth death process is in states $n~<~m$, where $m$ denotes the threshold. The second part enlists the surviving trajectories in which, at time $t$, the underlying birth death process is in states $n \geq m$. Needless to say, in the surviving trajectories, whenever the underlying birth death process is in states $n \geq m$, the sensor must be inactive. 

The possible trajectories in the first part are:
\begin{itemize}
    \item The birth death process does not reach the threshold (state $m$) until time $t$.
    \item The birth death process reaches state $m$ for the first time at time $t_1$ but the sensor is inactive at time $t_1$. The birth death process remains in states $\geq m$ for some time and the sensor remains inactive throughout. Then at time $t_2$, for the first time, the birth death process reaches state $m-1$ and then always stays below $m$ until time $t$.\\
    \vdots
\end{itemize}
We can sum up the probabilities of such trajectories as the following
\begin{align}
    \nonumber   S_{t}^{\textrm{(I)}}(a,n_0) &= \int_{t}^{\infty} dt' F_{t'}(m|n_0) \\
    \nonumber    & + \int_{0}^{t} \int_{0}^{t_2} \int_{t-t_{2}}^{\infty} dt_{3}\cdot dt_{2}\cdot dt_{1}~ F_{t_1}(m|n_0)\ p_{t_1}(i|a)\ F_{t_2-t_1}(m-1|m) \ e^{-\alpha (t_2 - t_1)}\  F_{t_3}(m|m-1)\\
    \nonumber    & + \int_{0}^{t} \int_{0}^{t_4} \int_{0}^{t_3} \int_{0}^{t_2} \int_{t-t_4}^{\infty} dt_5\cdot dt_4 \cdot dt_3 \cdot dt_2 \cdot dt_1 ~ F_{t_1}(m|n_0) p_{t_1}(i|a) F_{t_2-t_1}(m-1|m)\ e^{-\alpha (t_2 - t_1)}\\
    & ~~~~~F_{t_3 - t_2}(m|m-1) p_{t_3 - t_2}(i|i)F_{t_4 - t_3}(m-1|m) e^{-\alpha (t_4 - t_3)}~ F_{t_5}(m|m-1) + \cdots
\end{align}

Similarly, the possible trajectories in the second part are:
\begin{itemize}
    \item The birth death process reaches state $m$ for the first time at time $t_1$ and the sensor is inactive at $t_1$. Then, the birth death process remains in states greater than or equal to $m$ till time $t$ and the sensor too, remains inactive till time $t$.
    \item The birth death process reaches the state $m$ for the first time at time $t_1$ but the sensor is inactive at time $t_1$. The birth death process remains in states greater than equal to $m$ for some time and the sensor remains inactive throughout that time. Then at time $t_2$, for the first time, the birth death process comes below $m$ and stays below $m$ until time $t_3$. Next, at $t_3$, the birth death process reaches state $m$ for the first time after $t_2$ (while the sensor is inactive). Subsequently, the birth death process, till time $t$, remains in states higher than $m$ and sensor remains inactive.   \\
    \vdots
\end{itemize}

Summing up the probabilities of the aforementioned trajectories, we get
\begin{align}
\nonumber   S_{t}^{\textrm{(II)}}(a, n_0) &= \int_{0}^{t} \int_{t-t_1}^{\infty} dt_2 \cdot dt_1 F_{t_1}(m|n_0) p_{t_1}(i|a) e^{-\alpha (t-t_1)}  F_{t_2}(m-1|m) \\
\nonumber    & + \int_{0}^{t}\int_{0}^{t_3}\int_{0}^{t_2}\int_{t-t_3}^{\infty} dt_4 \cdot dt_3 \cdot dt_2 \cdot dt_1\ F_{t_1}(m|n_0)\ p_{t_1}(i|a)\ F_{t_2 - t_1}(m-1|m) e^{-\alpha (t_2 - t_1)}\\
 & ~~~~~ F_{t_3 - t_2}(m|m-1)\ p_{t_3 - t_2}(i|i)\ e^{-\alpha (t-t_3)} F_{t_4}(m-1|m) +\cdots
\end{align}
We can get the total survival probability as $ S_{t}(a, n_0) = S_{t}^{\textrm{(I)}}(a, n_0)+S_{t}^{\textrm{(II)}}(a, n_0)$. For brevity, it is convenient to define the following quantities
\begin{align}
 L(t)    &= \int_{t}^{\infty} dt' F_{t'}(m|n_0)\\
 g_{1}(t) &= F_t(m|n_0)\ p_t(i|a)\\
 g_{2}(t) &= F_t(m-1|m)\ e^{-\alpha t}\\
 g_{3}(t) &= F_t(m|m-1)\ p_t(i|i)\\
 g_{4}(t) &= \int_{t}^{\infty} dt' F_{t'}(m|m-1)\\
 g_{5}(t) &= e^{-\alpha t} \int_{t}^{\infty} dt' F_{t'}(m-1|m)
\end{align}

Eqs. (S15-S20) denote probability density functions for the following relevant quantities which have the following interpretations: (i) $L(t)$ denotes the probability that the birth death process does not reach $m$ till time $t$, given that initially it was in state $n_0 < m$; (ii) $g_1(t)$ is the probability that the birth death process reaches state $m$ for the first time at $t$, starting from an initial state $n_0$ and the sensor being inactive at time $t$ starting from an active state; (iii) $g_2(t)$ denotes the probability that the birth death process comes down to $m-1$ for the first time at $t$ starting from the state $m$ and the sensor remains inactive throughout the time $0$ to $t$, when the birth death process was in states $\geq m$;(iv) $g_3(t)$ denotes the probability that, starting from the state $m-1$, the birth death process reaches the state $m$ for the first time at $t$ and that the sensor is inactive at $t$ starting from the inactive state, (v) $g_4(t)$ is the probability that the birth death process does not reach the state $m$ till time $t$ starting from the state $m-1$, and finally (vi) $g_5(t)$ represents the probability that the birth death process remains in the states greater than or equal to $m$ till time $t$ with starting in the state $m$, and that the sensor remains inactive throughout this time.

With the above definitions in mind, summing Eqs.(S13) and (S14), we can write the probability that the threshold crossing event is not detected upto time $t$, with the state of the birth death process being $n_0$ and the sensor being active initially as
\begin{align}
\nonumber    S_{t}(a,n_0) &= L(t) + \int_{0}^{t} \int_{0}^{t_2} dt_{2} dt_{1}\ \ g_{1}(t_{1}) g_{2}(t_2 - t_1) g_{4}(t-t_2) +\\
\nonumber    & \int_{0}^{t} \int_{0}^{t_4} \int_{0}^{t_3} \int_{0}^{t_2} dt_4 dt_3 dt_2 dt_1\ \ g_{1}(t_1) g_{2}(t_2-t_1) g_{3}(t_3-t_2) g_{2}(t_4-t_3) g_{4}(t-t_4) +\cdots\\    & + \int_{0}^{t} dt_1\ g_{1}(t_1) g_{5}(t-t_1) + \int_{0}^{t}\int_{0}^{t_3}\int_{0}^{t_2} dt_3 dt_2 dt_1\ \ g_{1}(t_1) g_{2}(t_2-t_1) g_{3}(t_3-t_2) g_{5}(t-t_3)+ \cdots
\end{align}
The fact that all the terms form a convolution in time, the analysis is much simpler in Laplace space. We obtain the Laplace transformed survival probability as a sum of two infinite series
\begin{align}
  \tilde{S}_{s}(a,n_0) &= \tilde{L}(s) + \tilde{g}_{1}(s) \tilde{g}_{2}(s) \tilde{g}_{4}(s)+ \tilde{g}_{1}(s) \tilde{g}_{2}(s) \tilde{g}_{3}(s) \tilde{g}_{2}(s) \tilde{g}_{4}(s) + \cdots  + \tilde{g}_{1}(s) \tilde{g}_{5}(s)+ \tilde{g}_{1}(s) \tilde{g}_{2}(s) \tilde{g}_{3}(s) \tilde{g}_{5}(s)+ \cdots
\end{align}
where
\begin{align}
\tilde{L}(s) &= \frac{1~-~\tilde{F}_{s}(m|n_0)}{s} \\
\tilde{g}_{1}(s) &= \frac{\beta}{\alpha + \beta} \tilde{F}_{s}(m|n_0) - \frac{\beta}{\alpha + \beta} \tilde{F}_{s+\alpha+\beta}(m|n_0)\\
\tilde{g}_{2}(s) &= \tilde{F}_{s+\alpha}(m-1|m)\\
\tilde{g}_{3}(s) &= \frac{\beta}{\alpha + \beta} \tilde{F}_{s}(m|m-1) + \frac{\alpha}{\alpha + \beta} \tilde{F}_{s+\alpha+\beta}(m|m-1)\\
\tilde{g}_{4}(s) &= \frac{1-\tilde{F}_{s}(m|m-1)}{s} \\
\tilde{g}_{5}(s) &= \frac{1-\tilde{F}_{s+\alpha}(m-1|m)}{s+\alpha}~~.
\end{align}
Because of several repetitive factors in each term, we identify both the series as geometric series
\begin{align}
 \tilde{S}_{s}(a,n_0) = \tilde{L}(s) + \tilde{g}_{1}(s) \tilde{g}_{2}(s) \tilde{g}_{4}(s)\left[1+\tilde{g}_{2}(s) \tilde{g}_{3}(s) + \tilde{g}_{2}(s)^{2} \tilde{g}_{3}(s)^{2} +\cdots\right] + \tilde{g}_{1}(s) \tilde{g}_{5}(s)\left[1+\tilde{g}_{2}(s) \tilde{g}_{3}(s) + \tilde{g}_{2}(s)^{2} \tilde{g}_{3}(s)^{2} +\cdots\right]
\end{align}
both of which can be individually summed to give
\begin{align}
    \tilde{S}_{s}(a,n_0) &= \tilde{L}(s) + \tilde{g}_{1}(s) \tilde{g}_{2}(s) \tilde{g}_{4}(s)\left[\frac{1}{1-\tilde{g}_{2}(s) \tilde{g}_{3}(s)}\right]
     + \tilde{g}_{1}(s) \tilde{g}_{5}(s)\left[\frac{1}{1-\tilde{g}_{2}(s) \tilde{g}_{3}(s)}\right].
\end{align}
This allows us to write the Laplace transform of the survival probability as
\begin{align}
     \tilde{S}_{s}(a,n_0) &= \tilde{L}(s) + \frac{\tilde{g}_{1}(s) \tilde{g}_{2}(s) \tilde{g}_{4}(s)+\tilde{g}_{1}(s) \tilde{g}_{5}(s)}{1-\tilde{g}_{2}(s) \tilde{g}_{3}(s)}.
\end{align}
A term by term interpretation of the above equation is possible, as done for the Laplace transformed FDTD in the previous section.

In above analysis, we considered an initial state in which the sensor is active and the birth death process is in state $n_{0}(<m)$. If the initial conditions is such that the sensor is inactive, a similar analysis can be performed. But now the initial state of the birth-death process can be any number between $0$ to $N$. We summarise the results of such analysis below.
\begin{enumerate}
    \item $n_{0}<m$:
    \begin{align}
     \tilde{S}_{s}(i,n_0) &= \tilde{L}(s) + \frac{\tilde{g}_{6}(s) \tilde{g}_{2}(s) \tilde{g}_{4}(s)+\tilde{g}_{6}(s) \tilde{g}_{5}(s)}{1-\tilde{g}_{2}(s) \tilde{g}_{3}(s)}
    \end{align}
    with
    \begin{align}
        g_{6}(t)=F_t(m|n_{0})~p_t(i|i)~~~~&\text{and}~~~~\tilde{g}_{6}(s)= \frac{\beta}{\alpha + \beta} \tilde{F}_{s}(m|n_0) + \frac{\alpha}{\alpha + \beta} \tilde{F}_{s+\alpha+\beta}(m|n_0)
    \end{align}
$g_{6}(t)$ is the probability that starting with in the state $n_{0}$ with inactive sensor, the birth death process reaches the state $m$ for the first time at time $t$ and the sensor is inactive at $t$. $g_{6}(t)$ accounts for the probability of the initial part of the survival trajectories, which involve crossing of threshold state by the birth death process. Only this is the difference from initial active target case.
    
    \item $n_{0}\geq m$
    \begin{align}
     \tilde{S}_{s}(i,n_0) &= \tilde{L}_{1}(s) + \frac{\tilde{g}_{7}(s) \tilde{g}_{3}(s) \tilde{g}_{5}(s)+\tilde{g}_{7}(s) \tilde{g}_{4}(s)}{1-\tilde{g}_{2}(s) \tilde{g}_{3}(s)}
    \end{align}
    with
    \begin{align}
        L_{1}(t) = e^{-\alpha t}\int_{t}^{\infty} dt' F_{t'}(m-1|n_0) ~~~~&\text{and}~~~~ \tilde{L}_{1}(s) = \frac{1~-~\tilde{F}_{s+\alpha}(m-1|n_0)}{s+\alpha}\\
        g_{7}(t)=e^{-\alpha t}~F_t(m-1|n_{0}) ~~~~&\text{and}~~~~ \tilde{g}_{7}(s) = \tilde{F}_{s+\alpha}(m-1|n_{0})
    \end{align}
$L_{1}(t)$ is the probability that starting in the state $n_{0} \geq m$ with inactive sensor, the birth death process remains above the state $m-1$ and the sensor is inactive throughout this time $t$. $g_{7}(t)$ is the probability that starting in the state $n_{0}$ with the inactive sensor, the birth death process reaches the state $m-1$ for the first time at time $t$ and the sensor is inactive throughout this time. Similar to $g_{6}(t)$, probability $g_{7}(t)$ accounts for the initial part of the trajectories, which involve crossing of threshold state by the underlying birth death process. 
\end{enumerate}

\subsection{Splitting Probabilities}
A key feature of this modeling approach is that when the threshold crossing event is detected, the state of the birth death process is a random number which can take values lying in the set $\{m, m+1, \cdots, N \}$. A quantity of interest in such a scenario is the \emph{splitting probability} $H_{t}(\zeta)$ that gives us the probability that the birth death process is in state $\zeta$ while the threshold crossing event is detected. Using insights from the calculation of the survival probability, it is easy to obtain $H_{t}(\zeta)$.

It turns out that $\zeta=m$ is a special case. Detection of the threshold crossing event at the state $m$ can happen with the last jump of the birth death process being -- (1) from the state $m-1$ to $m$ where sensor can be active or inactive at the time of the jump or (2) from $m+1$ to $m$ where the sensor has to be inactive at the time of the jump (the becoming active after the jump while the birth death process is till in the state m). Whereas for $\zeta > m$, sensor has to be inactive at the time of the last jump, irrespective of the type of jump ($\zeta+1~to~\zeta~~or ~~\zeta-1~to~\zeta$). We consider $\zeta=~m$ and $\zeta>m$ cases separately.

We define a random variable $\tau_{m,m-1}$ which denotes the first time when the birth-death process reaches the state $(m-1)$ starting from the state $m$.

First we consider the case $k = m$. The trajectories leading to detection of the threshold crossing event with the birth death process in the state $m$ can be grouped in two parts -- (I) trajectories with the last jump of the birth death process at time $t$ from $m-1$ to $m$, (II) trajectories with the last jump of the birth death process from $m+1$ to $m$ at time $t$ and trajectories with the last jump of the birth death process from $m-1$ to $m$ but at some time $s<t$.

The possible trajectories in the part I are:
\begin{itemize}
    \item The birth death process reaches state $m$ for the first time at time $t$ and the sensor is active at the same time.
    \item The birth death process reaches state $m$ for the first time at time $t_1$ but the sensor is inactive. Then birth death process remains in states greater than $m$ till time $t_2$ and the sensor is inactive throughout. At time $t_2$, the birth death process reaches the state $m-1$. Then birth death process reaches the state $m$ at time $t$ for the first time after $t_2$ and the sensor is active at the same time.  \\
    \vdots
\end{itemize}
Summing up the probabilities of aforementioned trajectories gives
\begin{align}
\nonumber    H_{t}^{\textrm{(I)}}(m) &= F_t(m|n_0) p_t(a|a) \\
\nonumber    & + \int_{0}^{t}\int_{0}^{t_2} dt_1 F_{t_1}(m|n_0) p_{t_1}(i|a) F_{ t_2-t_1}(m-1|m) e^{-\alpha(t_2-t_1)} F_{t-t_2}(m|m-1) p_{t-t_2}(a|i)\\
\nonumber    & + \int_{0}^{t}\int_{0}^{t_4}\int_{0}^{t_3}\int_{0}^{t_2} dt_4 \cdot dt_3 \cdot dt_2 \cdot dt_1 F_{t_1}(m|n_0) p_{t_1}(i|a)F_{t_2-t_1}(m-1|m) e^{-\alpha(t_2-t_1)}\\
 & F_{t_3-t_2}(m|m-1) p_{t_3-t_2}(i|i) F_{ t_4-t_3}(m-1|m) e^{-\alpha(t_4-t_3)} F_{t-t_4}(m|m-1) p_{t-t_4}(a|i) + \cdots
\end{align}
The possible trajectories in the second part are:
\begin{itemize}
    \item The birth death process reaches the state $m$ for the first time at time $t_1$ and the sensor is inactive. Then birth death process remains in states greater than $m-1$ till time $t$ and the sensor is inactive throughout. At time $t$, the birth death process is in the state $m$ and at the same time sensor becomes active for the first time after $t_1$.
    \item The birth death process reaches the state $m$ for the first time at time $t_1$ and the sensor is inactive. Then birth death process remains in states greater than $m-1$ for some time and sensor is inactive throughout. At time $t_2$, the birth death process reaches the state $m-1$ for the first time after $t_1$. Then the birth death process reaches state $m$ at time $t_3$ for the first time after $t_2$ and the sensor is inactive. Then birth death process remains in states greater than $m-1$ till time $t$ and the sensor is inactive throughout. At time $t$, the birth death process is in state $m$ and at the same time sensor becomes active for the first time after $t_3$.\\
    \vdots
\end{itemize}
Summing up the probabilities of the aforementioned trajectories gives
\begin{align}
\nonumber  H_{t}^{\textrm{(II)}}(m) &= \int_{0}^{t} dt_1 F_{t_1}(m|n_0)p_{t_1}(i|a) P_{t-t_1}(m|m,\tau_{m,m-1}>t-t_1) \alpha e^{-\alpha t-t_1}\\
\nonumber & + \int_{0}^{t}\int_{0}^{t_3}\int_{0}^{t_2}dt_3 \cdot dt_2 \cdot dt_1 F_{t_1}(m|n_0) p_{t_1}(i|i) F_{t_2-t_1}(m-1|m) e^{-\alpha(t_2-t_1)} F_{t_3-t_2}(m|m-1)\\ & p_{t_3-t_2}(i|i) P_{t-t_3}(m|m,\tau_{m,m-1}>t-t_3) \alpha e^{-\alpha t-t_3} + \cdots
\end{align}
Then splitting probability for the detection at $m$ can be written as $H_{t}(m) = H_{t}^{\textrm{(I)}}(m) + H_{t}^{\textrm{(II)}}(m)$. For the brevity, we define the following quantities
\begin{align}
    h_{1}(t) &= F_t(m|n_0)\ p_t(a|a) \\
    h_{2}(t) &= F_t(m|m-1)\ p_t(a|i) \\
    h_{3}(k,m, t) &= \alpha e^{-\alpha t}\ P_{t}(k|m,\tau_{m,m-1}>t) ~~~ \text{For}~~~ k=~m,~m+1,~m+2,\cdots,~N 
\end{align}
We take the Laplace transform of $H_{t}(m)$, which gives

\begin{align}
 \tilde{H}_{s}(m) &= \tilde{h}_{1}(s)+  \tilde{g}_{1}(s) \tilde{g}_{2}(s) \tilde{h}_{2}(s)+ \tilde{g}_{1}(s) \tilde{g}_{2}(s) \tilde{g}_{3}(s) \tilde{g}_{2}(s) \tilde{h}_{2}(s)+ \cdots\\
& + \tilde{g}_{1}(s) \tilde{h}_{3}(m,m, s)+ \tilde{g}_{1}(s) \tilde{g}_{2}(s) \tilde{g}_{3}(s) \tilde{h}_{3}(m,m, s) + \cdots
\end{align}
where
\begin{align}
\tilde{h}_{1}(s) &= \frac{\alpha}{\beta+\alpha} \tilde{F}_s(m|n_0) + \frac{\beta}{\beta+\alpha} \tilde{F}_{s+\alpha+\beta}(m|n_0) \\
\tilde{h}_{2}(s) &= \frac{\alpha}{\beta+\alpha} \tilde{F}_s(m|m-1) + \frac{\alpha}{\beta+\alpha} \tilde{F}_{s+\alpha+\beta}(m|m-1)\\
\tilde{h}_{3}(m,m, s) &= \alpha ~\widetilde{LC}_{s+\alpha}(m) \\
\text{where}~~~LC_{t}(k)&=P_{t}(k|m,\tau_{m,m-1}>t) ~~~ \text{For}~~~ k=~m,~m+1,~m+2,\cdots,~N \nonumber
\end{align}
Factoring out the common terms, a geometric series can be identified.
\begin{align}
\nonumber    \tilde{H}_{s}(m) &= \tilde{h}_{1}(s)+ \tilde{g}_{1}(s) \tilde{g}_{2}(s) \tilde{h}_{2}(s) \left[1+ \tilde{g}_{2}(s) \tilde{g}_{3}(s)+  \tilde{g}_{2}(s)^{2} \tilde{g}_{3}(s)^{2}+ \cdots \right] \\
 & + \tilde{g}_{1}(s) \tilde{h}_{3}(m,m, s)\left[1 + \tilde{g}_{2}(s) \tilde{g}_{3}(s)+  \tilde{g}_{2}(s)^{2} \tilde{g}_{3}(s)^{2} + \cdots \right]
\end{align}
Both the geometric series can be summed up to give
\begin{align}
    \tilde{H}_{s}(m) &= \tilde{h}_{1}(s) + \frac{\tilde{g}_{1}(s) \tilde{g}_{2}(s) \tilde{h}_{2}(s)+ \tilde{g}_{1}(s) \tilde{h}_{3}(m,m, s)}{1- \tilde{g}_{2}(s) \tilde{g}_{3}(s)}.
\end{align}

Now we consider the case where $k= m+r$, where $r=~1,~2,~3,\cdots,~N-m$.  It is easy to see that $H_{t}(m+r)$ will be given by
\begin{align}
\nonumber    H_{t}(m+r) &= \int_{0}^{t} dt_1 F_{t_1}(m|n_0) p_{t_1}(i|a) P_{t-t_1}(m+r|m,\tau_{m,m-1}>t-t_1) \alpha e^{-\alpha t-t_1}\\
\nonumber & + \int_{0}^{t}\int_{0}^{t_3}\int_{0}^{t_2}dt_3 \cdot dt_2 \cdot dt_1 F_{t_1}(m|n_0) P_{t_1}(i|i) F_{t_2-t_1}(m-1|m) e^{-\alpha(t_2-t_1)} F_{t_3-t_2}(m|m-1)\\
& p_{t_3-t_2}(i|i) P_{t-t_3}(m+r|m,\tau_{m,m-1}>t-t_3) \alpha e^{-\alpha t-t_3}+ \cdots
\end{align}
Following similar steps as in the case of $k=~m$, we are led to the Laplace transform of $H_{t}(m+r)$ as
\begin{align}
   \tilde{H}_{s}(m+r) = \frac{\tilde{g}_{1}(s) \tilde{h}_{3}(m+r,m, s)}{1 - \tilde{g}_{2}(s) \tilde{g}_{3}(s)}
\end{align}

\subsection{SIS Model}
The susceptible-infected-susceptible (SIS) model of disease propagation is a stochastic model that describes the spread of a disease in a population, which we consider to be well-mixed. The population consists of two types of individuals, those that are susceptible to the infection, and those that are currently infected. The rate at which the disease is transmitted between individuals is $\lambda$, and the recovery rate for each infected individual is $\mu$. If $j$ represents the number of infected individuals, there are $j(N-j)$ pairwise contacts between infected and susceptible people in a well-mixed population. Each of the $j$ infected individuals recover at rate $\mu$. The rates of increase and decrease of the number of infected individuals is given by
 \begin{equation}
W_{+}(j) = \lambda j (N-j), \quad \quad W_-(j) = \mu j.
 \end{equation}
The parameter values chosen to obtain the curve for Fig.~5 from the main text are $N=15, m= 7, n_0 = 1, \lambda = 0.1$, and $\mu=0.4$.

 \subsection{Logistic Model}
The stochastic version of the logistic model describes the dynamics of a population of mortal agents, who can reproduce. A constant rate $B$ is assumed for each agent, which means that in a small time interval $dt$, each individual gives birth to a new individual with probability $Bdt$. For each agent, there is constant death rate (set to $1$), when the population size is low. However, for larger population sizes, the death rate increases by an amount which is quadratic in the size of the population. In the birth-death formulation, the transition rates are 
 \begin{equation}
 W_+(j) = B j, \quad \quad W_-(j) = j + K j^2/N,
 \end{equation}
 where $K$ determines the strength of influence of competition towards the death rates. The parameter values chosen to obtain the curve for Fig.~5 from the main text are $N=15, m= 8, n_0 = 4, B=1.5$, $K=0.1$, $\alpha=\beta=1$. 
 
 \subsection{Stochastic Resets}
 In the birth-death process with stochastic resets, apart from the simple birth-death dynamics, with a rate $r$, the underlying process can be \emph{reset} to the state $0$. This dynamics is reminiscent of fluctuating dynamics that undergo burst-like relaxations. Furthermore, one can also consider the process, where the reset happens to a dormant state, at rate $r$. When this reset happens, the underlying process spends some refractory time in this dormant state, and resumes its birth-death dynamics at a rate $g$. 
 
 
 
 The parameter values chosen to obtain the curve for Fig.~5 from the main text are $N=10, m=5, n_0 = 0, r=\alpha=\beta=1$ while the birth-death rates were chosen to be the same as the ones for the curves in Fig.~2. In the case where refractory period is also considered, we choose $g=1$.

\subsection{Derivation of the first passage time distribution conditioned on first detection at $T_d$}

We define $F_{T_f}(m|n_0,\sigma_0, T_d)$ to be the density that the first passage to threshold $m$ happens at time $T_f$, conditioned on the fact that the first detection of the threshold crossing event happens at $T_d$, and that the underlying process starts from a state $n_0$ whereas the sensor starts from state $\sigma_0$. As mentioned in the main text, for $T_f>T_d$, $F_{T_f}(m|n_0,\sigma_0, T_d)=0$.  For $T_f<T_d$, we are interested in the trajectories that reach the threshold $m$ for the first time at $T_f$, but go undetected, and eventually, the threshold crossing event is detected at time $T_d$. We can break each such trajectory in two parts: the evolution up to time $T_f$, and the evolution up to time $T_d$, starting from time $T_f$. The first part of each trajectory is a \emph{first passage trajectory}, {\it{i.e.}} one which reaches the threshold for the first time at time $T_f$. We can immediately also conclude that at time $t=T_f$, the state of the underlying birth death process must be equal to the threshold, and the state of the sensor must have been inactive. Furthermore, the second part of each trajectory is a \emph{first detection trajectory}, that starts from an initial state such that the birth death process is at the threshold and the sensor is inactive, and subsequently, the first detection of the threshold crossing event happens at time $T_d$. Putting this together, we have
\begin{equation}
F_{T_f}(m|n_0, \sigma_0,T_d) = F_{T_f}(m|n_0) \cdot p_{T_f}(i|\sigma_0)\frac{D_{T_d-T_f}(m,i)}{D_{T_d}(n_0,\sigma_0)}
\end{equation}
where the denominator $D_{T_d}(n_0,\sigma_0)$ is due to the fact that we are looking at the subset of trajectories, which are conditioned to undergo the first detection event at time $T_d$.

A similar argument enables us to see that for $T_f = T_d$, the probability $F_{T_f}(m|n_0, \sigma_0,T_d=T_f)= \frac{F_{T_d}(m|n_0)}{D_{T_d}(n_0,\sigma_0)}$.

\end{document}